# Propagation of ultra-high-energy cosmic rays in the magnetized cosmic web

**Jihyun Kim**[1]
*Osaka City University*
*Graduate School of Science, Osaka City University, Osaka, Osaka 558-8585, Japan*
*E-mail:* jhkim@sci.osaka-cu.ac.jp

**Dongsu Ryu**
*Ulsan National Institute of Science and Technology*
*Department of Physics, School of Natural Sciences, UNIST, Ulsan 44919, Korea*

**Soonyoung Roh**
*Ulsan National Institute of Science and Technology*
*Department of Physics, School of Natural Sciences, UNIST, Ulsan 44919, Korea*

**Jihoon Ha**
*Ulsan National Institute of Science and Technology*
*Department of Physics, School of Natural Sciences, UNIST, Ulsan 44919, Korea*

**Hyesung Kang**
*Pusan National University*
*Department of Earth Sciences, Pusan National University, Pusan 46241, Korea*

A high concentration of ultra-high-energy cosmic ray (UHECR) events, called a hotspot, was reported by the Telescope Array (TA) experiment, but its origin still remains unsolved. One of the obstacles is that there is no astronomical object, which could be the source, behind the TA hotpot. In an effort to understand the origin of the TA hotspot, we suggested a model based on the magnetized cosmic web structure. The UHECRs were produced from sources in the Virgo cluster and were initially confined by cluster magnetic fields for a certain period. Next, some of them preferentially escaped to and propagated along filaments. Eventually, they were scattered by filament magnetic fields, and come to us. To examine the model, we followed the propagation trajectories of UHE protons in a simulated universe with clusters, filaments, and voids, by employing a number of models for cosmic magnetic fields. In this study, we present some of the initial results, such as the ratio between the particles directly escaping from the clusters to the voids and particles escaping from the clusters to the filaments. We also discuss the feasibility of our model for the origin of the hotspot by examining the trajectories of the UHE protons.

---

[1]Speaker

1. **Introduction**

Since the existence of ultra-high-energy cosmic rays (UHECRs) was discovered more than 50 years ago, many observations have continued to increase event statistics in order to find out the nature and origin of UHECRs. Owing to these extraordinary efforts, some of the main questions on UHECRs have been resolved. It is confirmed that UHECRs originate from extragalactic sources because they cannot be confined within the galactic plane by the galactic magnetic field (GMF) [1]. It has been observed, in addition, that there is a sharp decrease of the UHECR energy spectrum above a threshold energy of around $5\times10^{19}$ eV, that is, the so-called Greisen-Zatsepin-Kuzmin (GZK) suppression [2, 3].

However, the origin of UHECRs has not been discovered yet. To find out the sources of UHECRs, there have been many correlation studies between the arrival direction distribution of UHECRs and the positions of astronomical source candidates, such as active galactic nuclei, radio galaxies, or starburst galaxies. However, these studies have not yet led to any conclusive results [4-11]. One of the main obstacles to narrowing down the sources of UHECR is that they are charged particles; thus, their propagation trajectories from the sources could be deflected by magnetic fields in the universe. The magnetic fields are ubiquitous in astrophysical environments at various scales from stars to clusters of galaxies. Therefore, investigations on the propagation of UHECRs in the magnetized large-scale structure (LSS) of the universe would play an important role in the search for the astronomical sources of UHECRs.

From this point of perspective, it is necessary to investigate the origin of a high concentration of UHECR events, the so-called TA hotspot, reported by the Telescope Array (TA) experiment [12]. Because there is no prominent source candidate behind the TA hotspot area, several astronomical objects that have large separation angles from the center of the hotspot, including M82, have been suggested as the source candidates of the hotspot events [13]. However, a definitive source remains inconclusive. This is because, to be classified as a source of the TA hotspot, those candidate objects require certain assumptions about the stronger strength of the GMF or the heavier mass composition of the UHECR as compared to the estimated values from the observational data. The primary particles of UHECRs are expected to be light nuclei from the TA experiment's data analysis [14]. In that case, a large angular distance between the position of M82 and the center of the TA hotspot, ~26.5º, indicates a significant deflection during the propagation; the model indicating that M82 is responsible for the TA hotspot requires a stronger strength of GMF than the typically known values of a few $\mu$G. In short, no plausible point source candidates have been identified in the direction of the TA hotspot, which may imply we need to pay attention not only to point sources, but also to the structures of galaxies in the local universe.

In the previous work [15], we focused on the local structures of galaxies and the propagation of UHECRs in the intergalactic magnetic fields (IGMF) of the local LSS. We reported the existence of filamentary structures of galaxies connected to the Virgo cluster and a statistically significant correlation of 5.1σ between the location of the galaxy filaments on the celestial sphere and the arrival directions of the TA events. To explain such a close correlation, we suggested a model based on the magnetized cosmic web structure that consists of galaxy clusters, galaxy filaments, and voids. It is known that magnetic fields exist in the medium of galaxy clusters, galaxy filaments, and even in the voids between them. The strengths of such magnetic fields are approximately a few $\mu$G in clusters, a few tens of nG in filaments, and are of an order of magnitude less than nG in voids. Their details are described in the previous work.

The suggested model for the origin of TA hotspot events is as follows. The UHECRs are produced at a source or sources in the Virgo cluster and confined by cluster magnetic fields for a period. Next, some of them preferentially escape to and propagate along filaments. Eventually, they are scattered by the turbulent magnetic fields in filaments and reach observers in the earth. Then, the arrival direction distribution of those UHECRs would be correlated with the positions of the filaments. It may appear as though there is an excess of events in a certain area, near which there are no other prominent UHECR sources, much like the appearance of the TA hotspot.

In this study, we examine the model by investigating the propagation trajectories of ultra-high-energy (UHE) protons in a simulated universe. In Section 2, we describe our model for the formation of LSS and the generation of magnetic fields, and the design the simulations of the propagation of the UHE protons. The results of the simulation of particle propagation will be provided in Section 3. Finally, we present our conclusions in Section 4.

## 2. Simulations

### 2.1 Model Universe Simulation

To study the propagation and scattering of UHECRs in the magnetized cosmic web, we produce a model universe through numerical simulation for the LSS formation using a particle-mesh/Eulerian cosmological hydrodynamics code [16]. The Lambda Cold Dark Matter ($\Lambda$CDM) cosmological model with the following parameters were assumed: baryon density of $\Omega_{BM} = 0.044$, dark matter density of $\Omega_{DM} = 0.236$, cosmological constant of $\Omega_\Lambda = 0.72$, Hubble parameter of $h \equiv H_0/(100$ km s$^{-1}$ Mpc$^{-1}) = 0.7$, RMS density fluctuation $\sigma_8 = 0.82$, and primordial spectral index n = 0.96. These parameters are consistent with the seven-year data from Wilkinson Microwave Anisotropy Probe observations [17].

A cubic box having a comoving size of 49 $h^{-1}$Mpc with periodic boundaries is generated using a $1440^3$ uniform grid zone. The resolution of the grid is 34.0 $h^{-1}$kpc, which is smaller than the gyro-radius of UHE protons in most zones. There are two clusters with X-ray weighted temperature T $\gtrsim$ 2.5 keV in the simulation volume. Since the temperature of the intracluster medium in the Virgo cluster is around 2.5 keV [18], they are appropriate for the simulation. We select one cluster with T = 2.8 keV as the sample cluster to produce UHE protons because the other cluster is a merging cluster that is not suited to mimic the Virgo cluster.

The generation of IGMF is seeded by the Biermann battery mechanism at cosmological shocks, and their evolution and amplification were followed passively along with flow motions [19]. We note that, alternatively, cosmological magnetohydrodynamic (MHD) simulations for the LSS formation can be used in UHECR propagation studies [20]; however, currently available numerical resources cannot reproduce the full development of MHD turbulence in the simulations of LSS formation. (See the detailed discussion in [21].) Therefore, we used the LSS formation simulation then included the IGMF passively.

The overall magnetic field strength is rescaled to reproduce the observed strengths. The magnetic field strengths are rescaled based on the magnetic field strength of the sample cluster core; the average strength of magnetic fields within 0.5 $h^{-1}$Mpc from the X-ray center of the cluster is used. In this work, we adopt two sets of model universes which are rescaled the cluster core region to 2 μG and 3 μG.

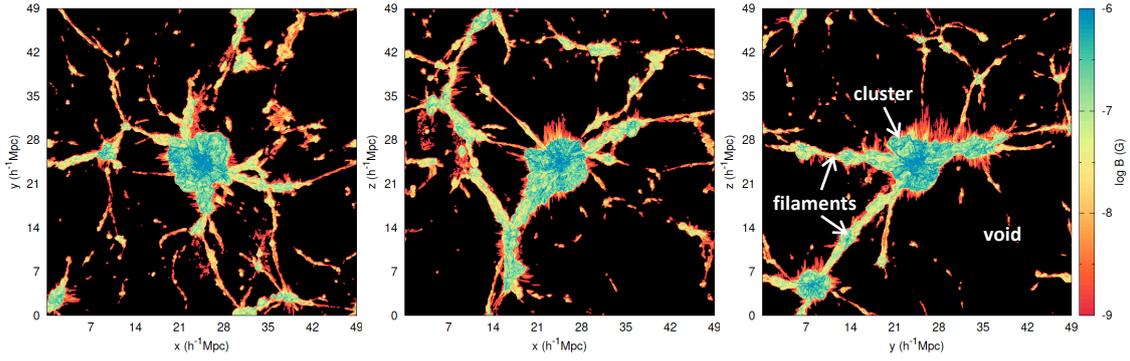

Figure 1. Two-dimensional slices of the magnetic field strength in the model universe. The sample cluster we used is located in the center of the box, and filamentary structures are seen around the cluster. The other spaces that are colored black, $B <$ nG, are void regions. The overall strength of the magnetic fields is rescaled to set the core region of the cluster to be ~2 μG.

## 2.2 Particle Trajectory Simulation

In the model universe described above, we inject UHE protons and examine their propagation trajectories assuming that their sources are in the core of the clusters. At random positions within the cluster core of 0.5 $h^{-1}$Mpc from the X-ray center position of the clusters, we launch $10^5$ protons having an energy of $6 \times 10^{19}$ eV in random directions. Next, we trace their trajectories with the relativistic equation of motions for charged particles under magnetic fields of $\boldsymbol{B}$ given by

$$\frac{d\boldsymbol{r}}{dt} = v$$

$$\frac{d\boldsymbol{v}}{dt} = \frac{Ze}{mc}(\boldsymbol{v} \times \boldsymbol{B}),$$

where $\boldsymbol{r}$, $\boldsymbol{v}$, and $m$ are the position, velocity, and mass of the charged particle, respectively. Here, UHE protons are assumed, thus $Z = 1$. In this study, the energy losses are ignored.

## 3.    Results

Figure 1 shows two-dimensional slices of the cosmic web structure of the model universe with magnetic field strengths. The color indicates the strength of the magnetic fields. The overall strength is rescaled to set the core region of the cluster to be ~2 μG. The sample cluster is located at the center of the box, and filamentary structures are seen around the cluster. Based on an examination of the shape of the sample cluster and considering the distribution of magnetic fields, its radius is defined as $R_{cluster}$ ~ 3.5 $h^{-1}$Mpc. In the core region, the average strength of the magnetic fields is ~2 μG (blue color), and it decreases toward the cluster outskirts. The average strengths of the magnetic fields are a few hundreds of nG (green and yellow-green mixed color) in the cluster outskirts, a few tens to hundreds of nG (yellow-green and yellow-orange mixed color) in the filaments, and smaller than nG (black color) in the voids, respectively.

It is expected that the size and magnetic field strength of the cosmic web affect the ability to confine the particle inside the structure. The gyro-radius of a charged particle is given by

$$r_g \sim 42\ h^{-1}\mathrm{kpc}\left(\frac{E/Z}{6\times 10^{19}\ \mathrm{eV}}\right)\left(\frac{B}{1\ \mu\mathrm{G}}\right)^{-1},$$

where $E$ and $Z$ are the particle's energy and charge. Considering the strengths of magnetic fields are around a few $\mu$G in clusters and a few tens of nG in filaments, the gyro-radius of a UHE proton having the same energy with the particle we used in this work would be a level of $h^{-1}$kpc in the cluster and would be an order of $h^{-1}$Mpc in the filaments. Whereas, the radius of typical clusters and the width of typical filaments are on the order of a few $h^{-1}$Mpc. Therefore, the UHE proton produced inside the cluster would be rather tightly confined by the cluster magnetic fields; it becomes easier to escape through the filaments because their widths and the gyro-radius of the particle in the filament are comparable.

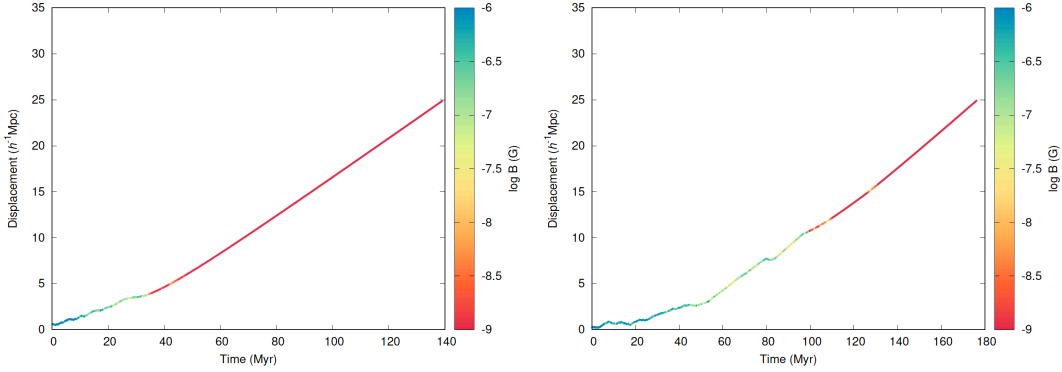

Figure 2. Displacement from the center of the cluster for two UHE protons. The left panel deals with the case of a particle that has directly escaped from the cluster to the voids, and the right panel is that of a particle that has escaped from the cluster to the filament, and has scattered to the voids.

In the particle trajectory simulation, it is observed that the turbulent magnetic fields in the cluster keep the UHE protons inside the cluster for a while. Next, as we expected, some particles escape from the cluster to the void region directly, while some particles propagate into the filaments. If we examine the trajectories of the UHE protons closely, their specific trajectories and last scattering points caused by the turbulent magnetic fields in the filaments are very diverse although they enter into the same filament. The specific trajectories are very sensitive to the configuration of magnetic fields.

Figure 2 shows displacements from the cluster center of two representative UHE protons as a function of time. The color bar depicts the magnetic field strength that the particles experience at their positions. The left panel is a typical example of particles that have directly escaped from the cluster to the voids. This UHE proton roams around the cluster for ~30 million years (Myr), then directly escapes to the voids where the strength of the magnetic field is weaker than $10^{-9}$ G. By contrast, the right panel shows how a particle escapes from the cluster via the filament. After wandering inside the cluster for more than ~50 Myr, this UHE proton propagates along the magnetic field line to the filament where the strength of the magnetic field is weaker than that of the cluster. It propagates around 10 $h^{-1}$Mpc along the filament and finally scatters to the voids.

Because we are interested in the propagation of UHECRs in the magnetized cosmic web to verify our model for explaining TA hotspot events, we focus on where they escape from and classify the particles into two categories according to their trajectories. Particles directly escaping from the cluster to the voids belong to one category, while particles escaping to the filaments, and then getting scattered to the voids belong to another. We examine the ratio between the particles directly escaping from clusters to the voids and particles escaping from the clusters to the filaments. In this work, we present the preliminary results. In the case of the 2 $\mu G$ rescaled cosmic web, the ratio is ~55% to ~45%; in the case of the 3 $\mu G$ rescaled cosmic web, the ratio is ~49% to ~51%. The results of this study confirm that it is possible for a UHE proton produced from a source in a galaxy cluster to escape through the galaxy filaments connected to the cluster.

## 4. Conclusion

We investigated the trajectories of UHECRs in the cosmic web produced through the numerical simulation of the LSS formation with IGMF seeded by the Biermann battery mechanism. The protons having energies of $6\times10^{19}$ eV were injected in the cluster core region to travel in random directions. We traced their trajectories and found that the number of particles directly escaping from the cluster and the number of particles escaping via filaments were similar, although their exact ratio depends on the magnetic field strengths of the cluster. The results support our model for the origin of the TA hotspot. This model predicts that UHECRs can be produced from a source or sources in the Virgo cluster and can escape through the connected filaments of galaxies.

This model can explain another characteristic distribution of TA UHECR data in the single frame. Observing the distribution of TA events [12], we can see a deficit of events toward the Virgo cluster, which is the closest to us (~16 Mpc) and contains ~1500 galaxies, composed of many active galactic nuclei, radio galaxies, and so on. Considering the fact that a UHECR flux would be inversely proportional to the square of the distance of a source, $f_s \sim 1/d^2$, it is expected that the distribution of UHECRs would be concentrated in the direction of the Virgo cluster. However, the observed distribution of UHECRs shows no excess UHECRs in the direction of the Virgo cluster. Interestingly, the Pierre Auger Observatory did not observe any excess of events around the Virgo cluster either [10]. We concluded that this discrepancy is one of the main reasons why there was no close correlation between UHECRs and active galactic nuclei in the previous studies [6, 7].

Based on the results of this study, almost half of the UHECRs propagate along the filaments of galaxies; thus, a lower number of observed events toward the cluster would be expected, which is logically linked to the excess of events observed toward the filaments. If the magnetic field line of the Virgo cluster is mostly closed in the direction to us, this distribution can be observed remarkably. Thus far, there is no direct observational evidence to reconstruct the exact configuration of the magnetic fields of the Virgo cluster; however, we introduce two intriguing observations that may hint at the topology of the field line in the Virgo cluster. To begin with, the brightest giant elliptical galaxies are aligned toward the base of the filaments and towards the hotspot. Second, the direction of the M87 jet expands towards the hotspot. Their projected axes are aligned similarly so that they are hard to distinguish. (See Fig. S3 in [15].) The peculiar structure of the magnetic field of the Virgo cluster, if it exists, might cause the characteristic distribution of UHECR events.

The results of this study confirm that a UHE proton produced from a source in a cluster of galaxies can escape to filaments of galaxies connected to the cluster and propagate along the filaments. Because the propagation trajectories of particles in the magnetized cosmic web depend on the specific configuration of magnetic fields, the information on the magnetic field distribution in the regions of the Virgo Cluster and the hotspot is required for realistic tests to reproduce the TA hotspot. Future astronomical projects for the exploration of intergalactic magnetic fields, such as the Square Kilometre Array, will make it possible for us to impose essential constraints on the magnetic fields in the cosmic web. More simulations with various IGMF models are in progress to overcome this limitation and to understand the propagation of UHECRs in the magnetized cosmic web.

## Acknowledgments

This work is supported by the Japan Society for the Promotion of Science (JSPS) through Grants-in-Aid for Scientific Research (S) 15H05741 and 19H05607; by the National Research Foundation (NRF) of Korea through grants 2016R1A5A1013277 and 2017R1A2A1A05071429.

## References

[1] A. Aab et al., *Observation of a large-scale anisotropy in the arrival directions of cosmic rays above $8\times10^{18}$ eV,* Science **357** (6357): p. 1266-1270 (2017)

[2] R.U. Abbasi et al., *Measurement of the flux of ultra high energy cosmic rays by the stereo technique,* Astroparticle Physics, **32** (1): p. 53-60 (2009)

[3] D. Ivanov, *Report of the Telescope Array - Pierre Auger Observatory working group on energy spectrum,* Proceedings of Science, PoS(ICRC2017)498

[4] R.U. Abbasi et al., *Search for cross-correlations of ultrahigh-energy cosmic rays with BL Lacertae objects,* Astrophysical Journal **636**(2): p. 680-684 (2006)

[5] J. Abraham et al., *Correlation of the highest-energy cosmic rays with nearby extragalactic objects,* Science **318**(5852): p. 938-943 (2007)

[6] H.B. Kim and J. Kim, *Statistical analysis of the correlation between active galactic nuclei and ultra-high energy cosmic rays,* JCAP03(2011)006

[7] H.B. Kim and J. Kim, *Revisit of correlation analysis between active galactic nuclei and ultra-high energy cosmic ray,* International Journal of Modern Physics D **22**(8) 1350045 (2013)

[8] T. Abu-Zayyad et al., *Correlations of the arrival directions of ultra-high energy cosmic rays with extragalactic objects as observed by the telescope array experiment,* Astrophysical Journal **777**(2): p. 88-95 (2013)

[9] A. Aab et al., *Searches for anisotropies in the arrival directions of the highest energy cosmic rays detected by the Pierre Auger Observatory,* Astrophysical Journal **804**(1): p. 15-32 (2015)

[10] A. Aab et al., *An indication of anisotropy in arrival directions of ultra-high-energy cosmic rays through comparison to the flux pattern of extragalactic gamma-ray sources,* Astrophysical Journal Letters **853**(2) L29 (2018)

[11] R.U. Abbasi et al., *Testing a reported correlation between arrival directions of ultra-high-energy cosmic rays and a flux pattern from nearby starburst galaxies using Telescope Array data,* Astrophysical Journal Letters **867**(2) L27 (2018)

[12] R.U. Abbasi et al., *Indications of intermediate-scale anisotropy of cosmic rays with energy greater than 57 EeV in the northern sky measured with the surface detector of the Telescope Array experiment,* Astrophysical Journal Letters **790**(2) L21 (2014)

[13] H.-N. He et al., *Monte Carlo bayesian search for the plausible source of the Telescope Array hotspot,* Physical Review D **93**(4) 043011 (2016)

[14] R.U. Abbasi, *Depth of ultra high energy cosmic ray induced air shower maxima measured by the Telescope Array Black Rock and Long Ridge FADC fluorescence detectors and surface array in hybrid mode,* Astrophysical Journal, **858**(2): p. 76-102 (2018)

[15] J. Kim et al., *Filaments of galaxies as a clue to the origin of ultrahigh-energy cosmic rays,* Science Advances **5**(1) eaau8227 (2019)

[16] D.S. Ryu et al., *A cosmological hydrodynamic code based on the total variation diminishing scheme,* Astrophysical Journal **414**(1): p. 1-19 (1993)

[17] E. Komatsu et al., *Seven-year Wilkinson Microwave Anisotropy Probe (WMAP) observations: cosmological interpretation,* Astrophysical Journal Supplement Series **192**(2) 18 (2011)

[18] R. Shibata et al., *Temperature map of the Virgo Cluster of galaxies observed with ASCA.* Astrophysical Journal **549**(1): p. 228-243 (2001)

[19] D.S. Ryu, H.S. Kang, and P.L. Biermann, *Cosmic magnetic fields in large scale filaments and sheets,* Astronomy & Astrophysics **335**(1): p. 19-25 (1998)

[20] S. Hackstein et al., *Simulations of ultra-high energy cosmic rays in the local Universe and the origin of cosmic magnetic fields,* Monthly Notices of the Royal Astronomical Society **475**: p. 2519-2529 (2018)

[21] R.M. Kulsrud et al., *The protogalactic origin for cosmic magnetic fields.* Astrophysical Journal, 1997. **480**(2): p. 481-491.